\documentclass[%
 reprint,
 amsmath,amssymb,
 aps,
]{revtex4-1}

\usepackage{graphicx}
\usepackage{dcolumn}
\usepackage{bm}
\usepackage{color}

\begin{document}


\title{Global scaling of the heat transport in fusion plasmas}


\author{Sara Moradi}
\email{sara.moradi@ukaea.uk}
\affiliation{Laboratory for Plasma Physics, LPP-ERM/KMS, Royal Military Academy, 1000-Brussels, Belgium}

\author{Johan Anderson}
\email{anderson.johan@gmail.com}
\affiliation{Department of Space, Earth and Environment, Chalmers University of Technology, SE-412 96 G\"{o}teborg, Sweden}

\author{Michele Romanelli}
\affiliation{Culham Centre for Fusion Energy, Culham Science Centre, Abingdon, OX14 3DB, UK}

\author{Hyun-Tae Kim}
\affiliation{EUROfusion Programme Management Unit, Culham Science Centre, Abingdon, OX14 3DB, UK}

\author{and JET contributors}
\email{See the author list of ``X. Litaudon et al 2017 Nucl. Fusion 57 102001".}
\affiliation{EUROfusion Consortium, JET, Culham Science Centre, Abingdon, OX14 3DB, UK}

\date{\today}

\begin{abstract}
A global heat flux model based on a fractional derivative of plasma pressure is proposed for the heat transport in fusion plasmas. The degree of the fractional derivative of the heat flux, $\alpha$, is defined through the power balance analysis of the steady state. The model was used to obtain the experimental values of $\alpha$ for a large database of the JET Carbon-wall as well as ITER Like-wall plasmas. The findings show that the average fractional degree of the heat flux over the database for electrons is $\alpha \sim 0.8$, suggesting a global scaling between the net heating and the pressure profile in the JET plasmas. The model is expected to provide an accurate and a simple description of heat transport that can be used in transport studies of fusion plasmas. 
\end{abstract}


\maketitle

\section{Introduction} Fusion plasmas are open systems with continuous energy input, inherently having a continuous drive of turbulence at many scales, i.e. similar to or approaching a scale free process, leading to a behaviour that is much more complex than standard diffusion. It is nowadays recognised that turbulence induced transport phenomena must be interpreted in the framework of the anomalous or turbulent diffusion as opposed to "normal" diffusion which is due to Brownian motion described by the Wiener process \cite{Wiener}. Anomalous transport is characterised by non-Gaussian (possess power-law tails) self-similar nature of the PDFs of particle displacement, and the anomalous scaling of the moments \cite{Anselmet1984,She1988,Castaing1990,Mordant2001,Wilczek2016,jimenez,Barenblatt,Frisch81}. 

Indeed, fluctuation measurements by Langmuir probes have provided abundant evidence to support the idea that density and potential fluctuations are distributed according to non-Gaussian PDFs and exhibit long-range correlations, see Refs. \cite{Moyer96,Jha97,Carreras96,Carreras98,Carreras99,Boedo2001,Sanchez2002,Zweben2002,LaBombard2002,antar,Terry2003,Lemoine,vanmiligan2}. Recent analysis of fluctuation measurements from Beam Emission Spectroscopy (BES) of MAST tokamak plasmas also shows evidence of skewed PDFs of density fluctuations in the near turbulence threshold regimes, see Ref. \cite{FoxPPCF2017,vanWykPPCF2017}. The skewed PDFs are suggested to be due to breaking of up/down and reflection symmetries of the fluctuation field by sheared flows \cite{ParraPOP2011}. Similar results were discussed in the Large Plasma Device facility at UCLA \cite{perez2006}, where vorticity probes (VP) were used to directly measurement the vorticity associated with $\mathbf{E}\times\mathbf{B}$ flow shear. These regimes possess complex dynamics and self-organisation properties that display uni-modal non-Gaussian features which is one of the signatures of intermittent turbulence with patchy spatial structure that is bursty in time \cite{pradalier2010,sanchez2008,negrete2005}. Therefore, the statistical properties of such dynamical chaotic systems fall outside the domain of the diffusive paradigm described by Brownian motion. 

A new school of thought based on fractional kinetics for systems with Hamiltonian chaos have gained momentum in different areas of applications, such as: particle dynamics in different potentials, particle advection in fluids, plasma physics and fusion devices, quantum optics, and many others \cite{Klafter1996,Metzler1993}. New characteristics of the kinetics are involved to fractional kinetics and the most important are anomalous transport, super-diffusion, weak mixing, and others. Fractional kinetics are tied closely to L\'{e}vy statistics, describing fractal processes (L\'{e}vy index $\alpha$ where $0 < \alpha \le2$ ) \cite{Levy}. L\'{e}vy statistics are considered to lie at the heart of complex processes such as anomalous diffusion that can be generated by random processes that are scale-invariant. From a physical point of view, these L\'{e}vy flights are the results of strong collisions between the particle and the surrounding environment, such as turbulent driven flows. The scale invariant and self similar nature of L\'{e}vy stable distributions gives rise to the occurrence of large increments of the velocity and position coordinates during small time increments, violating the local character of the collision integrals in the traditional deterministic equations. Experimental observation of the intermittent particle flux at the edge of ADITYA tokamak has been reported in Refs. \cite{JhaPRL1992, JhaPPCF1994,JhaPoP2003} where L\'{e}vy processes are thought to play key roles in the bursty fluctuations.

Even though the application of fractional kinetics in the study of turbulence phenomena shows great promise in resolving many open issues in the field, the current state-of-the-art has not gone beyond phenomenological levels such as Fractional Fokker-Plank equation (FFPE) \cite{smoradi2,smoradi3}, or L\'{e}vy random walk ideas \cite{smoradi1}. An important reason for this is the lack of a connection between the dynamic of the system to that of the L\'{e}vy index $\alpha$. Clearly a dynamical system moving through different phases, e.g. laminar to transitional to fully developed turbulent flow, can not be simply fixed by a given $\alpha$ \textit{a priori}. Instead it has to be linked to the underlying nonlinear dynamic of the system. A way to obtain the information regarding the value of $\alpha$ is to directly examine the experimental data.

In this work, we propose a novel global transport model based on a fractional approach. In this reduced model the aim is to construct a transport model that can represent most plasmas with high enough fidelity in terms of reproducing plasma profiles with significantly reduced computing resources. The main objective is to define the fractional index $\alpha$, of the heat flux in tokamak plasma experiments through a power balance analysis of the steady state profiles over the whole plasma region. Here, the divergence of the heat flux is modelled by a fractional derivative of the plasma pressure. The model depends on a single fractional index $\alpha$ that describe the degree of the global heat transport, i.e. the flux of the transported scalar at a point depends on the gradient of the scalar throughout the entire domain. This leads to constant heat diffusion coefficient to the cost of a fractional power exponent $\alpha$ over the radial profile in contrast to current modelling efforts where sharp variation in heat diffusivity while using a regular model with $\alpha=2.0$ in each radial point is found. Analysis show that the experimental values of $\alpha$ for a large database of the JET Carbon-wall as well as ITER Like-wall plasmas is $< 2$ with slightly lower values obtained for electrons than for ions. Here it is pertinent to keep in mind that the success of a fractional model indicates that there is a lack of physics in the current collisional and turbulent transport models which may be due to unphysical variations in the coefficient of heat diffusivity \cite{Romanelli98,Romanelli04,Romanelli10,smoradi14,kimNF2018,Rafiq2013,bourdelle2016}, namely the super-diffusive character of the heat transport. Note that although, we have employed the methodology for heat flux in magnetically confined plasmas, it is a general methodology that could be applied in any instance where a fractional model of dynamics is used.
\section{The global transport model}We start by examining the fluid equation for conservation of plasma energy in the following simplified form: 
\begin{equation}
\frac{3}{2}\frac{\partial }{\partial t} p_j(\mathbf{r},t)+ \nabla\cdot Q_j(\mathbf{r},t) =  H_j(\mathbf{r},t). \label{3}
\end{equation}
where $\mathbf{r}$ represents the cylindrical coordinate system $(R,Z,\phi)$ with $R$ being the radial position along the major radius, $Z$ being the vertical position, and $\phi$ being the toroidal angle. $Q_j$ describes the heat flux and $H_j$ is the net heating. The parallel (to the magnetic field lines) heat transport in tokamaks is significantly higher than the perpendicular one, and we can assume equilibration in parallel direction. Here therefore, we neglect the parallel heat flux and only consider the heat transport in $(R,Z)$ plain. In addition to Eq. (\ref{3}) an equation for the evolution of the density profile is also needed, however, we have limited our analysis to the heat transport. Note that in principle, a heat flux defined as (\ref{3}) includes all the processes that contribute to the steady state heat flux i.e. MHD, turbulence as well as the Neoclassical processes. We now introduce a modified equation including the following general form for the divergence of the heat flux (see Ref. \cite{Diego2010}):
\begin{equation}
\nabla\cdot Q_j(\mathbf{r},t) = D^{\alpha_{j}}_{|\mathbf{r}|}S_{j} p_j(\mathbf{r},t), \label{4}
\end{equation}
where $D^{\alpha_{j}}_{|\mathbf{r}|}$ is the fractional derivative operator with $\alpha_j$ as the index of the fractional derivative \cite{Diego2010}. To ensure the correct dimensionality, we have introduced $S_s$ as an effective (i.e. constant) super-diffusive transport coefficient with the dimensionality of $[L^{\alpha_j}/s]$. For $\alpha_j=2$, we get a purely diffusive model, and for $\alpha_j=1$ we obtain a purely convective transport model where the flux is defined as $Q_j = S_j p_j$, and $S_j \;[L/T]$ becomes the heat convective velocity. For $\alpha < 2$ therefore, the transport is so-called super-diffusive, and the lower the $\alpha_s$, the higher will be the level of the super-diffusive transport. 

To define $\alpha_s$, we propose to make use of the Fourier representation of (\ref{3}) as (see Ref. \cite{Diego2010}):
\begin{equation}
\frac{3}{2}\frac{\partial }{\partial t} \hat{p}_j(\mathbf{k},t) - |\mathbf{k}|^{\alpha_j} \hat{p}_j(\mathbf{k},t) =  \hat{H}_j(\mathbf{k},t). \label{5}
\end{equation}
Here, $\hat{X}\;$ represents the Fourier representation of quantity $X$, and $\mathbf{k} = \sqrt{k_R^2 + k_Z^2}$ where $k = (2\pi/L) [0\dots  M/2-1 \;\;0 \;\;-M/2+1\dots-1]$ with $L = 2m$ in the radial direction and $L=4m$ in the vertical direction. $M=256$ modes have been considered. For simplicity we have assumed $S_j=1$. This means that all the physics contributing to the transport namely collisional, neoclassical and turbulence processes, is contained within the fractional index $\alpha_j$. Through a power balance analysis using Eq. (\ref{5}), we can find the following expression for $\alpha_j$: 
\begin{equation}
\alpha_j = \frac{\log(\frac{\hat{H}_j - (3/2)\partial_t \hat{p}_j}{-\hat{p}_j})}{\log|\mathbf{k}|}. \label{6}
\end{equation}
The fractional index, $\alpha_j$, as defined above, will be a complex number and a function of $\mathbf{k}$ (see for example the computed values of $\alpha(k_R,k_Z)$ in Fig. \ref{f2} (a,b)). The imaginary part and the variations in $\mathbf{k}$ represent the detail structures in the profiles of heating and the pressure e.g. an off or on axis heating schemes, or presence of edge/core transport barriers. Here, we are interested in a fractional transport model with a constant fractional degree, and therefore, in order to define a scale independent fractional index we perform an averaging over the scales. The detail information encapsulated within the $\mathbf{k}$ dependence of $\alpha$'s should in principle be included in the super-diffusive transport coefficient, $S$. However we restrict our fractional model to one specific scale dependence which, as an average, is expected to be the dominant scale dependence (i.e. constant $\alpha$), and it will be suitable as a fast transport model. We note that by taking constant $\alpha < 2$ the dependence on different scales is closer to the experimental spectrum as compared to a regular diffusion model. Furthermore, with decreasing $\alpha$ the dampening of small scale modes such as Electron Temperature Gradient (ETG) modes will be less prominent, due to the reduced dissipation effect of the term $|\mathbf{k}|^{\alpha}$. Hence, such modes may be of enhanced importance in plasmas with a strong non-diffusive component.

For the steady state, the time derivative term in the numerator vanishes, and the value of $\alpha$ depends on the ratio of the heating power to the pressure. In a sense, the single value of $\alpha$ obtained in this way, gives the relation between the two profiles of pressure, and the net heating deposition at the steady state. It represents the final relaxation state of the pressure profile due to the all of the different turbulent mechanisms, and collisions that move the energy and particles in and out of the confined plasma region. Note, that due to the central limit theorem, a combination of Gaussian and L\'{e}vy processes will not result in a Gaussian process, therefore finding a fractional index $\alpha\ne2$ indicates that there are contributions from non-gaussian processes which resulted in such a fractal scaling.


In the following the results of our analysis for a selected database of the JET tokamak plasmas are presented. 
\section{The JET dataset and the results} The analysis is performed using a large dataset from the JETPEAK database \cite{HenriNF2017} of the JET Carbon (C) and ITER Like Wall (ILW) experiments. The analysed dataset contains 1256 samples from 868 different plasma shots. Each sample is an average over a stationary state for $~1s$, therefore, the time derivative of the pressure in the relation (\ref{5}) is neglected. Moreover due to the time averaging, an average effect from transitional MHD behaviours such as Edge Localised Modes (ELMs), and Sawteeth crashes are accounted for within the analysed profiles.    

$T_i = T_e$ is assumed but where CX spectroscopy data is available, measured Ti is used. $100\%$ Carbon and Beryllium as the only present impurity in the C-wall and ILW plasmas respectively. To compute the ion density, in the C-wall plasmas measured effective charges are used. In ILW uniform $Z_{eff} = 1.2$ is assumed. For electrons the net heating is computed following $= H_{in}-H_{Rad} -H_{ie}$, and for the ions the net heating is computed as $= H_{in}+ H_{ie}$. Note that $H_{ie}=0$ when $T_i = T_e$. The input heating profiles, i.e. $H_{in}$, are obtained from beam deposition code PENCIL \cite{PENCIL} and for ICRH by the code PION \cite{PION}. $H_{Rad} = 20\%H_e$ is assumed.

\begin{figure}[h]
\centering
\includegraphics[width = 0.4\textwidth]{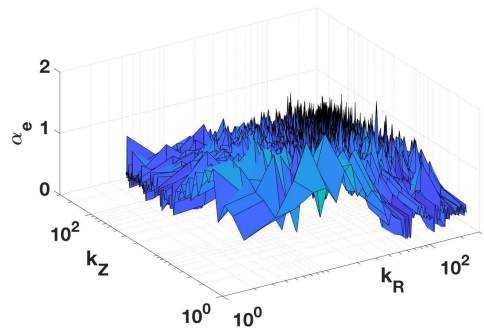}
\caption{\label{f2} The computed value of $\alpha_e$ as function of $k_R$ and $k_Z$ for the plasma shot $\# 58158$, the corresponding averaged value is $\alpha_e= 1.0$. This discharge is an ELMy H-mode pulse with regular Type I ELMs during the steady state phase.}
\end{figure}

\begin{figure}[h]
\centering
\includegraphics[width = 0.52\textwidth]{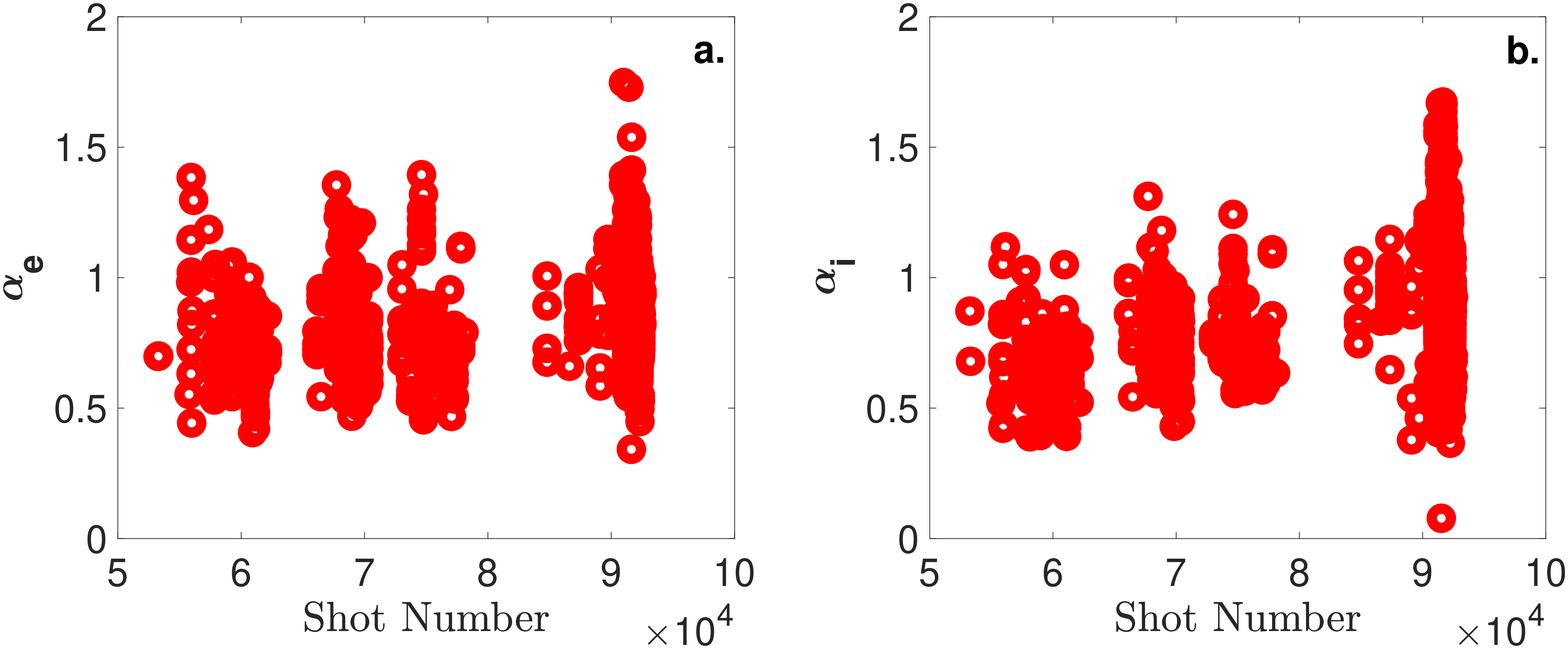}\\\includegraphics[width = 0.52\textwidth]{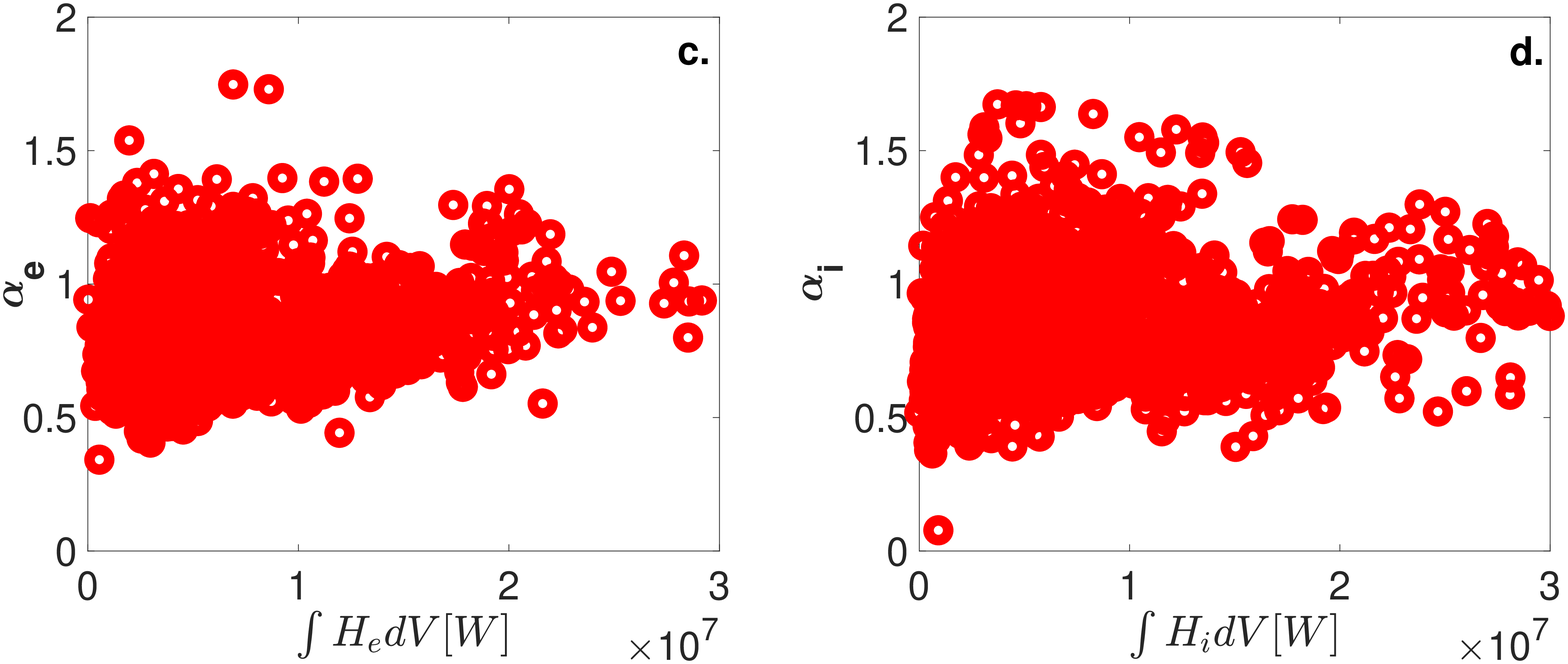}
\caption{\label{f5} The computed $\alpha_e$ (a,b) and $\alpha_i$ (c,d) as functions of the plasma shot number, and the volume integrated net heating power for the selected JETPEAK dataset.}
\end{figure}

\begin{figure}[h]
\centering
\includegraphics[width = 0.34\textwidth]{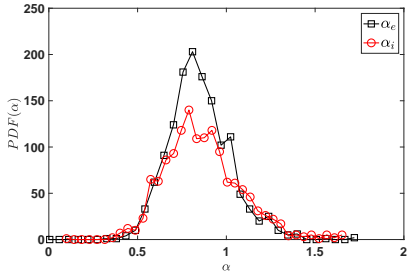}
\caption{\label{f5.1} The histogram of the computed $\alpha_{e}$ (black line with square symbols), and $\alpha_{i}$ (red line with circle symbols) for the selected JETPEAK dataset.}
\end{figure}

The degree of the globality of the transport processes were determined by computing $\alpha_{e,i}$'s following the relation (\ref{6}). An example of the computed $\alpha_{e}$ as function of mode numbers, $k_R$ and $k_Z$, for the plasma shot $\#58158$ is shown in Fig. \ref{f2}. To test the accuracy of the Fourier space derivatives for the heating and pressure profiles, the first and second derivatives were computed both in the real and the Fourier spaces where good agreements were found. The value of fractional index for the first derivative is found $=1$  and for the second derivative $=2$, as expected. However, at higher mode numbers, the values suffer from numerical errors and thus, a high-$k$ cut off (cut off point is $|k_{R,Z}| > 60$) is applied before averaging over $k^{>}_{R,Z}$. 
Figures \ref{f5} (a-d) show the values of $\alpha_{e,i}$'s as functions of plasma shot numbers (a,b) where the shot numbers $< 80000$ and $> 80000$ are for the C-wall, and the ILW plasmas, respectively. Figures \ref{f5} (c,d) show the $\alpha_{e,i}$'s as functions of volume integrated net heating. As can be seen here, in all of the considered plasmas and for both electrons and ions, the computed fractional degrees are less than $2$. The nature of the heat transport in these plasmas, therefore, is expected to obey a non-diffusive model. The values of $\alpha_{e,i}$ cover a wide range from $\sim 0.5$ and $\sim1.5$ due to the wide differences in the heating, fuelling and scenario schemes across these plasmas. However, a general convergence trend towards $\alpha_{e,i}\sim 1$ is observed with an increase in the total power (see Figs. \ref{f5} (c,d)). Figure \ref{f5.1} shows the histograms of the fractional index $\alpha_{e}$ (black line with square symbols), and $\alpha_i$ (red line with circle symbols). The peak of the distributions are around $\alpha_{e,i} \approx 0.8$ and the standard deviations are $STD_{\alpha_e} = 0.17$, and $STD_{\alpha_i} = 0.21$.   

\begin{figure}[h]
\centering
\includegraphics[width = 0.35\textwidth]{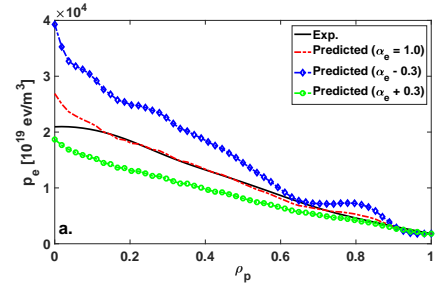}
\caption{\label{f7.1} Comparison of the experimental (black solid line) electron pressure profile vs normalised poloidal flux index $\rho_p$, and the predicted profile following the global transport model in (\ref{4.1}) (red dashed-dotted line) for the plasma discharges $\# 58158$. The predicted pressure profiles with $\pm 0.3$ above (blue dotted line with diamond symbols) and below (green dotted line with circle symbols) the computed values of $\alpha_{e}$ are also shown. This discharge is an ELMy H-mode pulse with regular Type I ELMs during the steady state phase.}
\end{figure}

In the steady state, the relation (\ref{5}), can be used to predict the pressure profiles from the heat deposition profile, and $\alpha$ following the expression:
\begin{equation}
p_{e,i} (\mathbf{r}) = IFT[-|\mathbf{k}|^{-\alpha_{e,i}}\hat{H}_{e,i}(\mathbf{k})], \label{4.1}
\end{equation}
where $IFT$ represents inverse Fourier Transformation. Figure \ref{f7.1} shows the predicted $p_e$ for the plasma discharge $\#58158$. The experimental profile (black solid line) is compared to the profile predicted by using the computed $\alpha$ (dashed-dotted red line). The predicted pressure profiles with $\pm 0.3$ above (blue dotted line with diamond symbols) and below (green dotted line with circle symbols) the computed value of $\alpha_{e}$ are also shown. As can be seen in Fig. \ref{f7.1}, the best agreement between the experimental and the predicted pressure profiles is found for the computed value of $\alpha_e$.  
\section{Fidelity of the model} 
\begin{figure}[h]
\centering
\includegraphics[width = 0.5\textwidth]{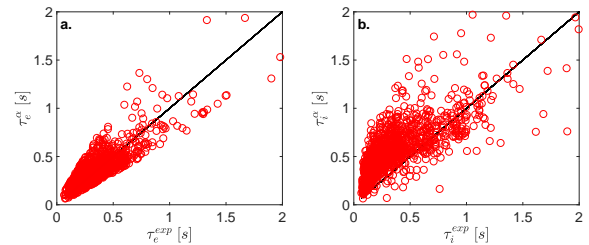}
\caption{\label{f9} The electron (a) and ion (b) energy confinement times, $\tau^{Exp}$, computed from experimental pressure profiles as a function of the energy confinement time, $\tau^{\alpha}$, computed from the predicted pressure profiles following the relation \ref{4.1}, are shown.} 
\end{figure}

As a measure of fidelity of the global model (\ref{4}), we have compared the ion and electron energy confinement times obtained from the experimental pressure profiles with the predicted ones following the expression: $\tau = \int p_{e,i} dV/\int H_{e,i} dV$, where $H_{e,i}$ are the experimental heat deposition profiles. Figure \ref{f9} (ab) shows the experimental $\tau_{Exp}$ as a function of predicted $\tau_{\alpha}$. Here, the confinement times were computed by applying the volume integration over the whole plasma region from the core to the last closed flux surface. A good agreement is found for the electron energy confinement times. The agreement for the ions is less good with the predicted profiles mostly overestimated as compared to the experimental values within the JETPEAK dataset. However due to the absence of $T_i$ measurements in many of the cases, the predictions for the $p_i$ profiles are limited.
\section{Discussion and Conclusion} A global heat flux model based on a fractional derivative of plasma pressure is proposed for the heat transport in the fusion plasmas. The degree of the globality of the heat transport is defined through the power balance analysis. In the proposed fractional model, a single constant fractionality index, $\alpha$, is used as the dominant global scale dependence of the transport which is modified as compared to a diffusive model where $\alpha=2$. Our aim with this work is to find a minimalistic (i.e. with the least amount of parameters involved) transport model that can predict most plasmas, therefore ignoring the detail nature and the classifications of the transport processes involved, and bundle their average (time/radial) effect into one constant parameter, $\alpha$. 

The method was used to study the heat transport in a selected set of JET plasmas, including C-Wall and ITER like Wall, L-mode, H-mode,
with many different heating and fuelling schemes from a wide range of experimental programs and plasmas with and without ELMs and various MHD modes active. The average fractional degree of the heat flux over the dataset was found as $\alpha \sim 0.8$. These results suggest that a global profile dependency between the net heating and the pressure profile in the JET plasmas exists which results in the relaxation of the pressure profiles to that of the heating deposition profile with a global decay rate, i.e. $|\mathbf{k}|^{-\alpha}$. Thus the profiles from the database of JET stationary phase are consistent with a constant fractional index $\alpha \approx 0.8$ on average, if one assumes a constant diffusivity profile and equal for all of the cases. Using the assumption of a universal transport coefficient the actual behaviour of the turbulent transport processes correspond to an $\alpha$ index significantly lower than $2$ on average. The 0-D model was then used to predict the pressure profiles, and the comparison between the energy confinement time obtained from the experimental and the predicted kinetic profiles show a very good agreement specially for the electrons. In the future, the proposed fractional transport model could be used as a feedback control of the plasma stability and control in real time by predicting profiles and thus providing a tool to detect and perhaps prevent or mitigate destructive transport events. It should be noted, that in some cases there is a wider range of $\alpha$ parameters over the database in particular because these plasmas are in essence very different with one another on many factors such as the NBI or ICRH input power, fuelling scheme, ELM control, etc. What we are
observing however, is that a significant number of these plasmas fall into a similar range for alpha parameter specially as the input power is increased yielding a transport model with predictive power in a wide parameter regime.

The most common transport models e.g. TRANSP, JETTO and ETS, assume locality of transport resulting in the diffusive approach. Experimentally it has been found that at least two major observations do not agree with this assumption: 1) the predicted heat diffusivity coefficient in most cases does not provide the observed level of transport, 2) global interplay between the core, edge and SOL shows features of non-locality of transport such as long range correlations. Indeed, it is widely accepted that turbulent transport does not completely follow a locality law, in the sense that the relation between the flux and the gradient is not precisely linear, thus naturally yielding a fractional index differing from $2$.

Finally, we would like to make a note that this study is the first of its kind and its findings are expected to encourage further discussion on the validity and the mathematical limitations of our current models to address global properties of transport in fusion plasmas. Future work includes implementing the present transport model as an option the TRANSP framwork in order for the model to more widely tested.

\begin{acknowledgments}
We thank Dr. Tariq Rafiq, Henri Weison and the JET task force leaders for sharing their pearls of wisdom with us during the course of this research, and for their comments on an earlier version of the manuscript, although any errors are our own and should not tarnish the reputations of these esteemed persons. This work has been carried out within the framework of the EUROfusion Consortium and has received funding from the Euratom research and training programme 2014-2018 under grant agreement No 633053. The views and opinions expressed herein do not necessarily reflect those of the European Commission.
\end{acknowledgments}



\end{document}